# Atomic Calligraphy:

# The Direct Writing of Nanoscale Structures using MEMS

Matthias Imboden[1], Han Han[2], Jackson Chang[1], Flavio Pardo[3], Cristian A. Bolle[3], Evan Lowell[4], David J. Bishop[1,2,5*]

[1]Electrical & Computer Engineering, Boston University, 8 Saint Mary's Street, Boston MA 02215, USA

[2]Department of Physics, Boston University, 590 Commonwealth Avenue, Boston MA 02215, USA

[3]SLIM Line, Bell Labs, Alcatel-Lucent, Murray Hill, NJ, USA

[4]Department of Mechanical Engineering, Boston University, 110 Cummington Mall, Boston, MA 02215

[5]Division of Materials Science and Engineering, Boston University, 15 Saint Mary's Street Brookline, MA 02446

[*] djb1@bu.edu

**We present a micro-electromechanical system (MEMS) based method for the resist free patterning of nano-structures. Using a focused ion beam (FIB) to customize larger MEMS machines, we fabricate apertures as small as 50 nm on plates that can be moved with nanometer precision over an area greater than 20x20 μm$^2$. Depositing thermally evaporated gold atoms though the apertures while moving the plate results in the deposition of nanoscale metal patterns. Adding a shutter only microns above the aperture, enables high speed control of not only where but also when atoms are deposited. Using a shutter, different sized apertures can be selectively opened and closed for nano-structure fabrication with features ranging from nano- to micrometers in scale. The ability to evaporate materials with high precision, and thereby fabricate circuits and structures in situ, enables new kinds of experiments based on the interactions of a small number of atoms and eventually even single atoms.**

In conventional lithography methods a large amount of material is deposited or grown, and subsequently the desired geometry is etched out. This method no longer works when the number of atoms wanted in a device is reduced to the level of single or few atoms. This is certainly true for optical and e-beam lithography[1, 2]. The MEMS devices we present here allow full patterning control of the amount and location of a desired material. It is argued that combining a nanoscale aperture with a high speed MEMS shutter enables single atom placement. Without the need for resists and etchants, and the ability to deposit in situ, many materials chemically incompatible with lithography can be patterned in a bottom up method. A typical example would be lithium or other alkali metals as well as reactive elements like sulfur.

Other resist free, direct nano-patterning techniques exist. Dip Pen nanolithography is a technique based on using an AFM probe to flow liquid 'inks' onto a substrate[3] and oxidation nano lithography that alters the local chemistry [4-6]. Both methods are impressive, though limited by the types of compatible deposition materials and operation environments. Stencils can be used for resist free deposition[7, 8]. In such methods the patterned geometry is an exact copy of the stencil. Most similar to our approach, apertures or stencils can also be placed on an AFM cantilever and used for resist free patterning of arbitrary shapes[9, 10]. These systems lack the integrated shutter and speed possible with MEMS devices. Their setup and control is also significantly more complex, making it difficult to integrate these techniques with other technologies.

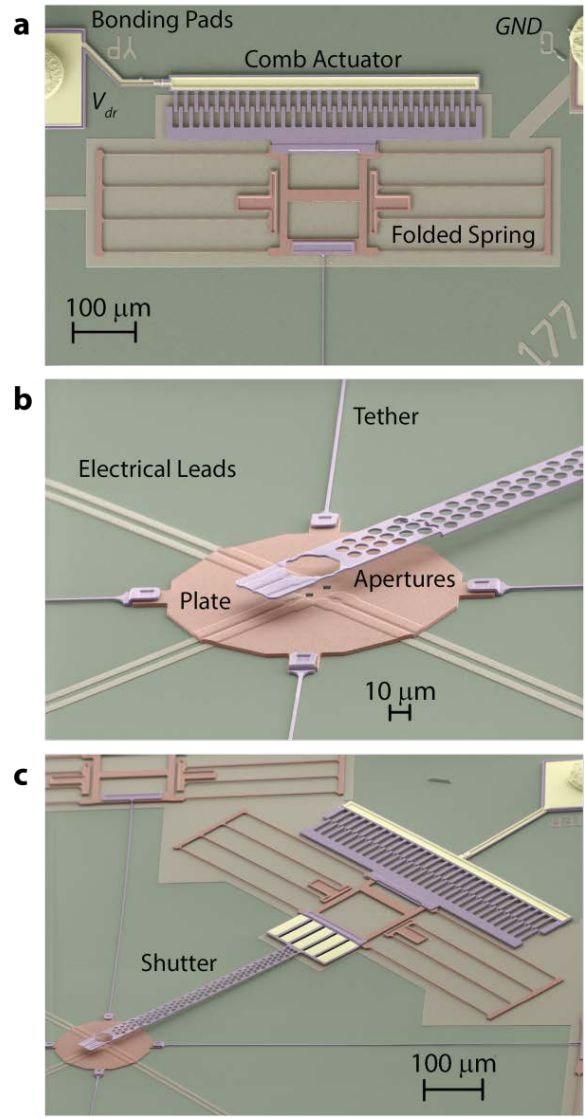

**Figure 1 | False color SEM micrographs of the MEMS writer. a**, Electrostatic comb actuator ($F = 0.924 \times V^2$ [nN]) and folded springs ($k = 0.344$ N/m). Preset leads enable electrical access. **b**, Plate with apertures. Aperture diameter ranges from 0.05-2 μm. Tethers connect the plate to the folded springs. **c**, Double plate design. Top plate functions as a high speed shutter.

MEMS devices have been successfully demonstrated in a wide range of operating conditions. They are robust, can operate at cryogenic temperatures and are highly reliable[11]. In situ deposition can be used for ultra-clean quench condensed films made of reactive elements and compounds[12]. Using a relatively simple control circuit, electrostatic actuation allows for high displacement resolution and negligible power consumption. The MEMS writers themselves are manufactured in a foundry at low cost using industry standard lithography methods in a scalable way. Because of their low cost and batch fabrication, our approach allows for the scalable manufacturing of large numbers of nano-devices. The chips are single use devices enabling high flexibility and turnover. They are simple to operate and inexpensive so non-experts can rapidly fabricate their devices to study novel materials or NEMS structures.

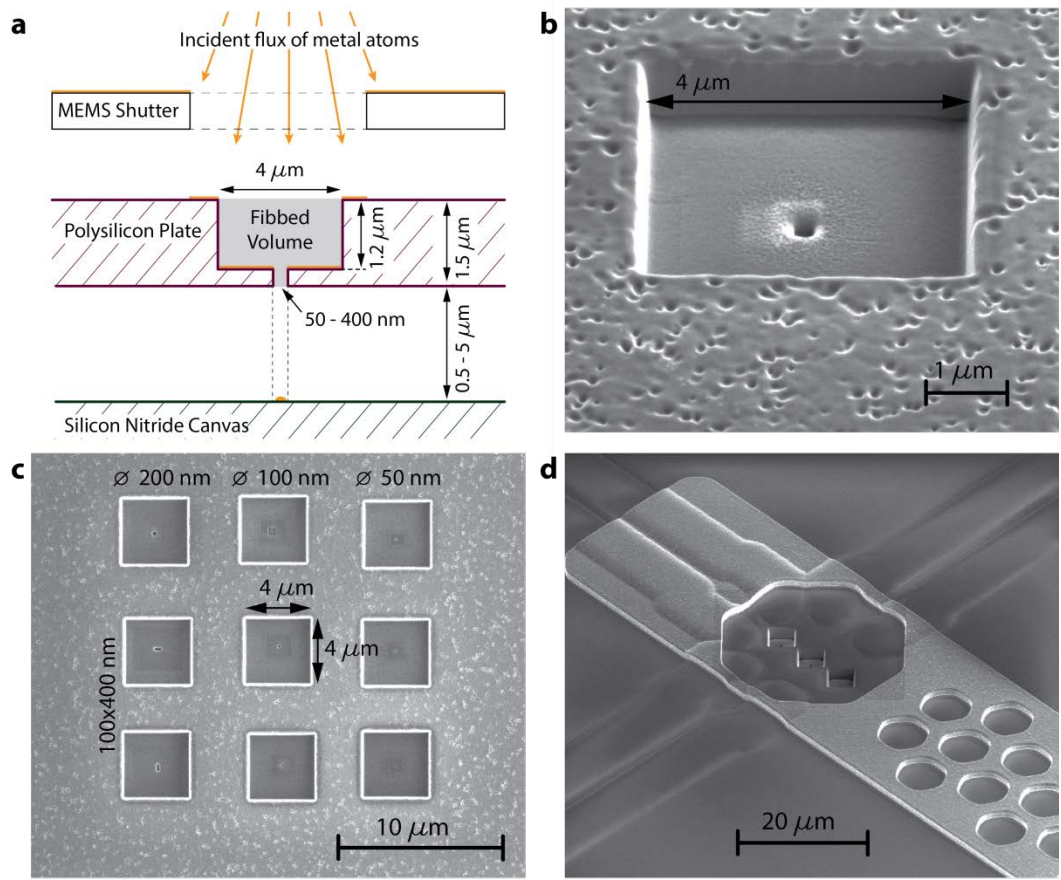

**Figure 2 | Aperture fabrication. a**, Diagram of milled aperture, including source, shutter and silicon nitride canvas. **b**, Two step aperture. Initial 4 x 4 x 1.2 µm³ trough thins and smoothens the polysilicon plate. High precision mill pokes through to make the nm scale aperture. **c**, Array of apertures with various sizes and shapes. **d**, Pre-release plate, shutter and apertures.

Our devices are fabricated using the MEMSCAP PolyMUMPs process[13], described in the supplementary materials (supplementary Fig. S1). Such optical lithography based structures can be manufactured economically at a large scale with ~1 μm resolution. The individual writers, consisting of electrostatic comb actuators, folded springs and a central plate, sit on a 2.5 x 2.5 mm$^2$ die stack of silicon nitride. Figure 1 depicts micrographs of the structure and its components. The smallest feature or dot that can be patterned is defined by the aperture dimension. We use a FIB to mill an aperture in the plate of the writer as illustrated in Figure 2. By leveraging the strengths of the scalable PolyMUMPs process with the nano-scale resolution of a FIB we obtain MEMS devices 2 mm across with customized feature sizes below 50 nm.

MEMS technology is expanding at a tremendous rate. Current systems include gyroscopes, mass sensors, accelerometers, pressure sensors, switches, time standards and more[14, 15,16]. . Due to the large linear response regime of poly-silicon, there is practically no mechanical fatigue, enabling a high cycling number and rate with no hysteretic effects. The small size scale results in a high mechanical frequency and quality factor. Therefore the writers are, to a large extent, mechanically isolated from their surroundings and external vibrations, making them easier to use in a wide range of applications.

The central plate is suspended over the substrate by four doubly folded flexure springs[17] and tethers. The springs and tethers can be combined into a single device that can move laterally >10 μm in all four quadrants[18], using the substrate as an additional electrode enables z-axis pull in[18, 19] (Supplementary Fig. S7). The device is actuated by four electrostatic comb drives, each attached to a folded spring. See Figs. S2-S5 in the supplementary material for a full discussion of the electromechanical response of the device. The writing occurs in a thermal evaporator (see supplementary Fig. S6). Gold is deposited through the apertures as the MEMS writer is actuated. The process is described in the methods section.

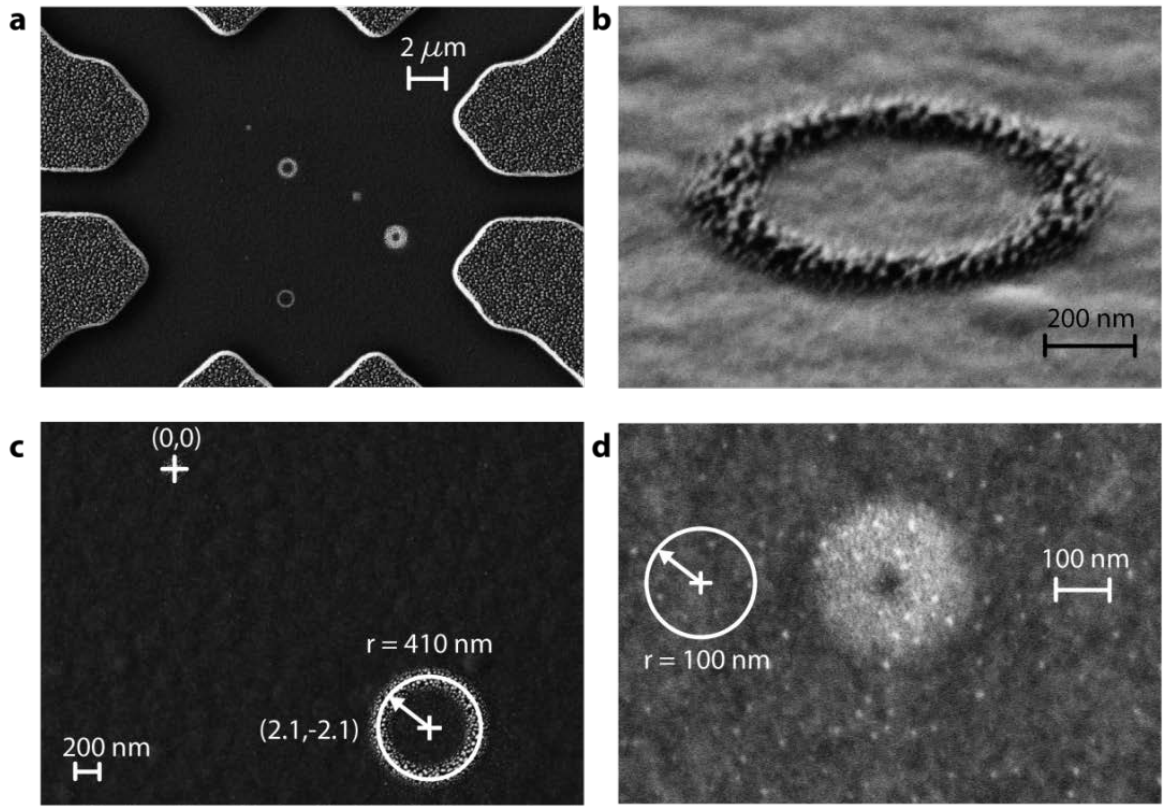

**Figure 3 | Demonstration of deposited rings of gold. a**, A plate with three apertures of varying sizes (100, 200 and 400 nm) illustrates rings deposited off center, where the dots indicated the zero voltage position. **b**, Close-up of 400 nm radius ring of the 100 nm aperture structure, positioned at (2.1,-2.1) µm. **c**, Top down view of the same structure as in (b) including the (0,0) position. **d**, Shown is a 100 nm radius ring with an aperture of 50 nm, resulting in a line width of 88 nm.

Figures 3 through 6 demonstrate the technological capabilities of the MEMS device. Figure 3 depicts rings traced out by different sized apertures. The smallest structure shown is a ring with a radius of 100 nm and a line width of just under 90 nm (Fig. 4), drawn with a 50 nm aperture. Figures 3a and 3b exhibit a ring drawn off center with the zero voltage location marked by a dot resulting from the stationary aperture. These images illustrate the high level of control and reproducibility. Even after 10 passes the rings are traced out identically over the first pass. In addition to mechanical vibration, the fuzziness of the edge is the result of geometric effects due to the height of the aperture over the substrate (Supplementary Fig. S7). This sets a limit to the resolution of the device and can be improved by pulling the plate towards the nitride. Figure 4 depicts an array of nine 4.76 x 2.84 µm² infinity symbols patterned in parallel with differently shaped apertures. All apertures are on the same plate so the deposition occurs in parallel. A larger array printed from 4x4 identically milled apertures is shown in the Fig. S9 of the supplementary materials. Figure 4b illustrates how tilting the aperture results in different line widths and thicknesses depending on the direction in which the plate moves. The resulting trace looks like a Moebius loop written by atomic calligraphy, analogous to the nib of a fountain pen. Such apertures can be used selectively to pattern narrow or broad structures by choosing the direction of motion.

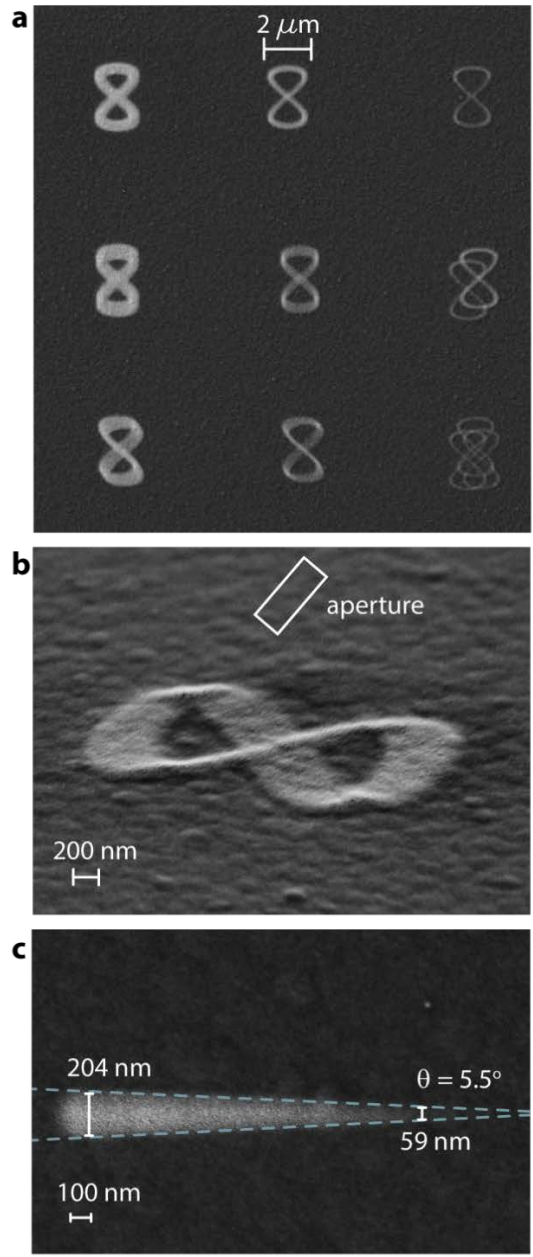

**Figure 4 | Results: Arrays, Moebius loops and line sharpening**. **a**, Array of apertures of varying sizes and shapes. **b**, Tilted image of gold deposited through a tilted aperture results in an image of a Moebius Loop. Using this calligraphy method the width of the line depends both on the shape of the aperture and its orientation relative to the direction of the motion of the plate. **c,** Aperture thinning. As gold is deposited the aperture fills in. This process can be used to controllably narrow apertures of varying sizes. Depositing 260 nm of gold results in a narrowing of the trace by ~225 nm. An initially 204 nm diameter line is closed completely with deposition still visible below 60 nm. The rate of closing and minimum feature discernible is defined by the material used as well as the source-aperture-canvas geometry (see supplementary Fig. S7b).

The FIB dictates the smallest aperture that can be manufactured. However, it is possible to further narrow the aperture by selectively allowing it to fill in with the deposition metal, as illustrated in Figure 4c. During the deposition the aperture was moved in a straight line. As the hole fills in the trace becomes narrower. This process is analogous to the 'ion-beam sculpting' used for fabrication molecular sized nanopores[20]. The narrowest line eventually blurs out and is no longer discernible. The lower limit of the line width is dictated by height of the aperture above the nitride as well as the level of mechanical vibration. For this case, using gold as a deposition material, the aperture diameter shrinks by ~0.87 times the total deposition thickness. The pore fills in much faster than would be expected by purely geometric considerations as has been reported in stencil masks[8]. It is suggested that surface phenomenon result in the movement of the gold clusters 'adatoms' that are subsequently trapped in the hole, which is thereby closed. It is expected that the deposition material, deposition rate, source-MEMS geometry and plate temperature all contribute to the rate of closing. Where the closing of the aperture sets an upper limit on the total amount of material that can be pattered, it also allows for the controlled narrowing of the aperture. To reach the molecular scale a feedback mechanism must be implemented[20]. In a device printing process, the narrowing would be the penultimate step, hence in situ customizing the aperture before the smallest feature is patterned. In operation, we envision using an array of holes, writing larger structures first with larger apertures. The smaller holes are kept clean with the shutter until needed.

In order to control, at high temporal and spatial precision, when an aperture is open or shut we have integrated an on chip MEMS shutter within a few microns over the holes. The open and closed positions of the shutter are depicted in Figure 5a and 5b respectively. This corresponds to comb potentials of 40 and 70 V. To demonstrate the ability to open and close a single aperture in a two aperture plate, a set of concentric circles are drawn with 400 and 800 radii. The shutter is set to open and close so that each circle is truncated into four quadrants. The resulting pattern for the open aperture that is always open is depicted in Figure 5c. The corresponding deposition through the second aperture, which is selectively closed, is illustrated in Figure 5d. Figure 5e is a SEM micrograph of the adjacent patterns. These depositions resulted from 11 passes, illustrating the repeated opening and closing of the shutter a total of 44 times. Though here the speed of the shutter is not demonstrated, in principle the rate is determined by the resonant response, resulting in an open-closed cycle of ~100 μs period. Methods to reduce this timescale are discussed in the supplementary material illustrated by Fig. S10.

The resolution of the MEMS writer is determined by the electronics control circuit, as well as the electrical and mechanical noise. As the displacement resulting from the applied voltage is given by $x = \frac{1}{2k}\frac{dC}{dx}V^2$ the displacement sensitivity with respect to the voltage is $dx = \frac{1}{k}\frac{dC}{dx}VdV = 2.54 \times 10^{-3}\frac{\mu m}{V^2}VdV$. This means that at 1 V actuation and a voltage noise as high as 1 V would result in only 2.54 nm displacement noise. At 100V actuation, corresponding to a displacement of $x = 12.7\ \mu m$, and a voltage noise of 10 mV the resulting displacement noise is still only 2.54 nm, well below our detection resolution at this time. The high quality factor and resonance frequency isolates the mechanical mode from external vibration noise which typically falls off as *1/f*. Current estimates based on measurements indicate typical vibrations are on the order of 40 - 60 nm. Pulling the central plate down to the substrate sharpens the deposition. The primary effect of the pull down is eliminating any geometric smearing due to the finite source size (Supplementary Fig. S7). Furthermore, any mechanical motion due to electrical or vibrational noise is eliminated once the plate makes contact with the substrate and can no longer move independently. Surface diffusion, that clogs the aperture, may also broaden the features[8]. Adding full control of the plate height will enable intentional focusing and defocusing depending on the application.

Dynamic smearing will occur due to the ring down time of the structure (supplementary Fig. S11). This can be compensated using slow small step sizes and active feedback control[21].

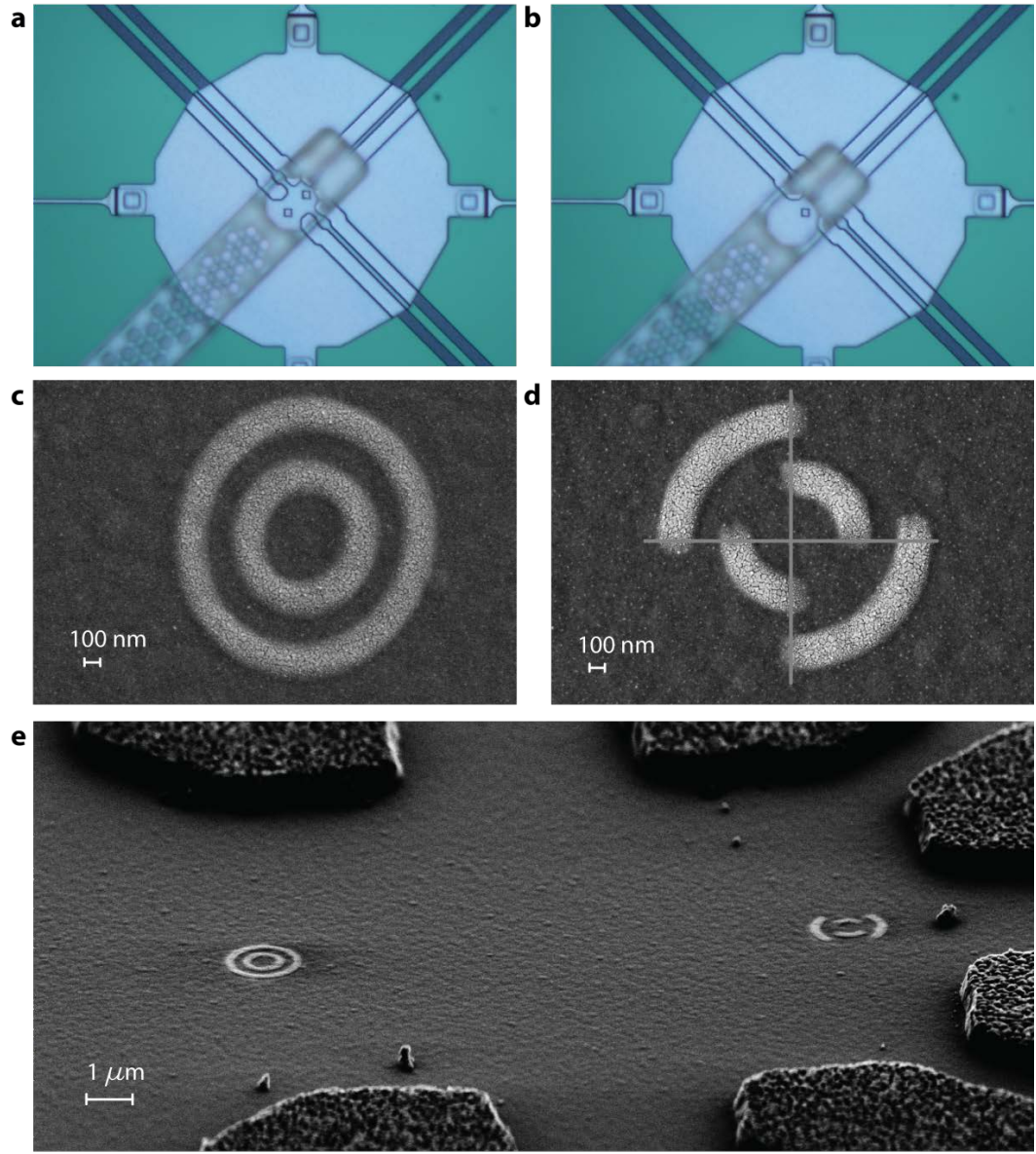

**Figure 5 | Plate and shutter device.** Optical image of **a**, open shutter ($V_{SH}$ = 40 V) with two apertures visible, and **b**, closed shutter ($V_{SH}$ = 70 V) with one aperture visible. **c**, Concentric circles drawn by a continuously open aperture. **d**, Truncated concentric circles resulting from the opening and closing aperture. **e**, Concentric circles side by side illustrate MEMS shutter functionality.

Considering the small apertures, low evaporation rates and high shutter speeds, we predict it is possible to allow stochastically a single atom to pass through the aperture. Evaporating with a deposition rate of one monolayer per second, means approximately 34,300 gold atoms are passing through a 50 x 50 nm$^2$ aperture every second. Remembering that the shutter can open and close in under 100 µs, it is estimated that just over 3 atoms can pass per cycle. Using smaller apertures, lowering the deposition rate and driving the shutter above its resonance speed are all possible and will push the average number of atoms passing though the aperture well below order one (See supplementary information and Fig. S10 for estimates of aperture-shutter response times).

In summary, we present a novel MEMS based shadow writer. By customizing optical lithography based MEMS devices with a focused ion beam, we mill apertures in a plate that can be controlled dynamically over a 20 x 20 µm$^2$ area. Using this device we pattern structures with feature sizes < 50 nm. As the patterning method described here is the final fabrication step, this technology can be used to build electronic circuits and structures in situ out of materials typically not used in nanolithography. A MEMS shutter makes it possible to stochastically control the number of atoms passing though the aperture down to of order one. The small footprint of this MEMS device, as well as the thermal and mechanical properties of polysilicon, allow this technology to be implemented directly in a cryogenic environment. Integrating this writer together with a MEMS based evaporator, as well as resonant sensors for deposition rate and temperature creates a cheap and versatile 'fab on a chip'. This will enable new mesoscopic experiments of quench condensed films, quantum dots and single atom effects.

**Methods**

**Aperture milling using a focused ion beam.** This is accomplished by a two-step process illustrated in Fig. 2. The thinner the plate, the smaller the aperture can be made (as a high aspect ratio reduces the milling resolution). For this reason initially a 4x4 µm$^2$ large trough is milled most of the way though the plate. This also smoothens the polysilicon. As a result, it is now possible to poke an aperture as small as 50 nm through the remaining 200-300 nm of silicon. In principle this process could be optimized to produce apertures in diameter just above the gallium ion beam size, which can be as small as ~5-10 nm[22, 23]. Even smaller pinholes (possibly down to ~10 nm) can be manufactured using a TEM provided the silicon is sufficiently thinned[24]. Currently the smallest aperture used resulted in an evaporated pixel size of ~42 x 57 nm$^2$ as depicted in supplementary Fig. S8.

**Deposition while actuating the MEMS**. The MEMS device is packaged and ball bonded into an 8 pin DIP chip holder and covered with a lid containing a ~75 µm hole centered over the plate. Patterning is conducted in a thermal evaporator at pressures in the mid 10$^{-6}$ Torr range. Feedthroughs enable electrical access to the writer. In a typical deposition, a crucible is heated resistively until a stable gold flux of 0.20-0.25 nm/s is obtained. The desired pattern is converted into the corresponding voltage coordinates, where for each axis a separate voltage source us used. Whenever actuating in one direction the voltage on the opposite capacitor is set to zero. Depending on the desired pattern a single pass or multiple passes are performed during a 10 – 20 minute period. Where the total deposition may be as high as 260 nm the pattern itself is typically less than 10 nm thick. A macroscopic shutter protects the MEMS during the ramping of the source. Some devices include a MEMS shutter only microns over the apertures enabling high speed opening and closing of individual apertures on a single plate. . Essentially, any compound that can be evaporated in vacuum could serve as a source material, especially if only a small number of atoms are desired.

**Instrumentation**. SEM images were taken with a Zeiss Supra 55 microscope. The FIB used was a FEI Quanta 3D FEG.

**Acknowledgements**

The authors thank Vladimir Aksyuk for help with the MEMS designs and discussions on the electromechanical response of these systems and Gregory McMahon for technical assistance with the FIB used to fabricate the apertures. This research is funded in part by Boston University.

**Author Contributions**

The experiments were designed by M.I. and D.B. The MEMS devices were designed by M.I. F.P and C.B. M.I. milled the structures with the FIB. H.H., J.C and M.I. conducted the experiments. E.L. and M.I. performed the finite element simulations. M.I. and D.B analyzed the results and wrote the manuscript. All authors discussed the findings and commented on the manuscript.

**Additional information**

Supplementary information is available in the online version of the paper. Reprints and permission information is available online at http://www.nature.com/reprints. Correspondence and requests for materials should be addressed to D.B.

**Competing Financial Interests statement**

The authors declare no competing financial interests.

# Atomic Calligraphy:

# The Direct Writing of Nanoscale Structures using MEMS

# (Online Supplementary Information)

Matthias Imboden, Han Han, Jackson Chang, Flavio Pardo, Cristian A. Bolle, Evan Lowell, David J. Bishop

Contents

1. *MEMS Fabrication using the PolyMUMPs Process*

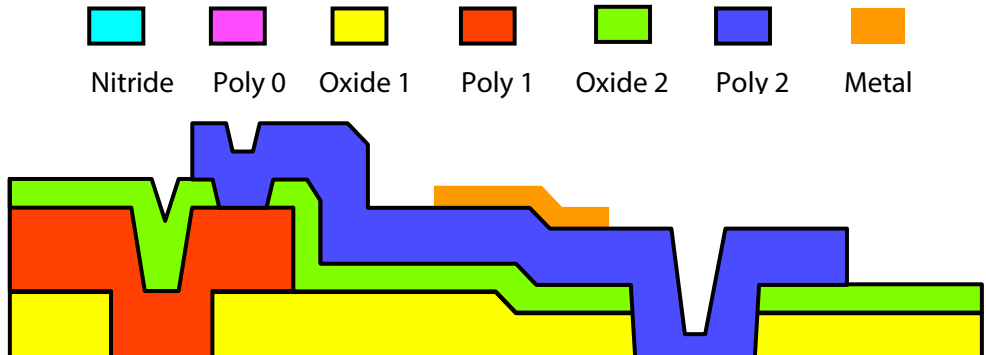

**Figure S-1. PolyMUMPs layers**[1]

Figure S1 depicts the PolyMUMPs layers consisting of nitride, silicon oxide, polysilicon and metal. The PolyMUMPs fabrication steps result in three doped polysilicon device layers, two of which can be suspended. The comb actuators are a double layer 3.5 μm thick, the springs are made of the 2 μm thick polysilicon layer and the plate and four tethers are fabricated from the top 1.5 μm thick polysilicon device layer. Table S1 contains the corresponding material properties as reported by MEMSCAP.

The sacrificial oxide layers define the vertical height of the polysilicon; compressive strain however can considerably shift the equilibrium position, as is described in section 3.

**Table S-1. PolyMUMPs Material Properties**[1] Density = 2320 kg m$^{-3}$, Young's Modulus = 158 ± 10 GPa, Poisson's Ratio 0.22 ± 0.01, Fracture Strength 1.21 ± 0.8 to 1.65 ± 0.28 GPa, size dependent.

| Run #100 150mm wafers | | | | | |
|---|---|---|---|---|---|
|  | **Thickness (A)** | **Standard Deviation (A)** | **Sheet Resistance (ohm/sq)** | **Resistivity (ohm-cm)** | **Stress (MPa)** |
| **Nitride** | 5,879 | 276.31 |  |  | 88 T |
| **Poly0** | 5,053 | 11.964763 | 26.07666667 | 1.32E-03 | 13 C |
| **Oxide1** | 19,895 | 660.47154 |  |  |  |
| **Poly1** | 19,988 | 96.927522 | 12.24 | 2.45E-03 | 3 C |
| **Oxide2** | 7,624 | 236.40869 |  |  |  |
| **Poly2** | 15,242 | 108.09713 | 21.10333333 | 3.22E-03 | 9 C |
| **Metal** | 5,467 |  | 0.0555 |  | 10 T |

## 2. Comsol Finite Element Simulation of Doubly Folded Springs, Tethers and Comb Drive

The central plate (150x150x1.5 µm³) is suspended over the substrate by four doubly folded flexure springs (with arms 2 µm thick and wide and 260 µm long) and tethers (1.5 µm thick, 2 µm wide and 509 µm long). Using elasticity theory and finite element simulation we determine the spring constant along the axis of actuation to be 0.703 N/m. The corresponding mode frequency is ~10 kHz. The springs are stiff in the perpendicular axis, but the tethers are not. This means that the four springs and tethers can be combined into a single device that can move laterally >10 µm in all four quadrants, using the substrate as an additional electrode enables z-axis pull in[2, 3].

### 2.1 Mechanical Properties of Folded Flexure Springs and Tethers

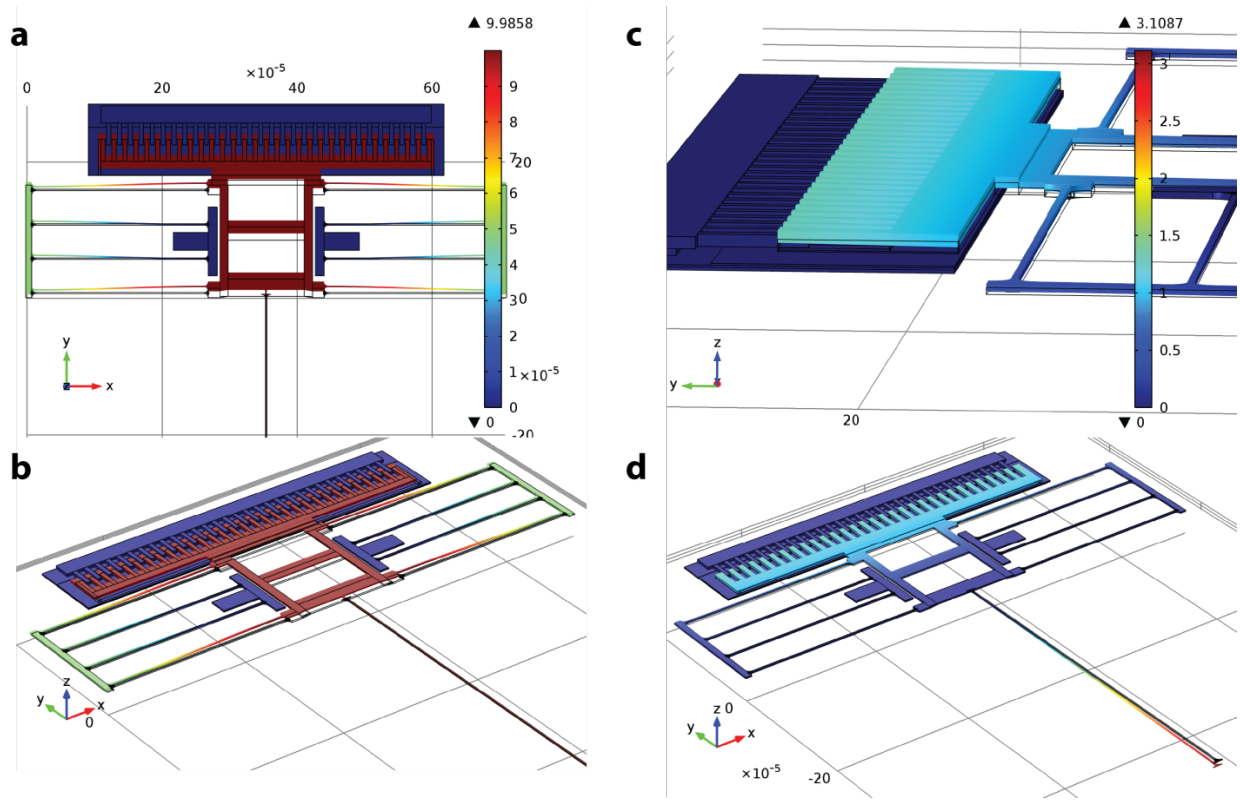

**Figure S-2. Actuated Folded Springs and Tether. a**, top view of vertical displacement of 10.0 µm due to a 7.03 x10⁻⁶ N force. **b**, tilted view of a**. Finite element simulation of out of plane forces. c**, side view of out-of-plane displacement due to 0.12 x 10⁻⁶ N force acting on combs. **d**, tilted view of c. The comb only rises by 1.3 µm, but the tether is lowered by 3.1 µm.

**Table S-2. Mechanical parameters of springs and tethers.**

| Measure | Value | Units |
|---|---|---|
| $E$ | 160 | GPa |
| $L$ | 255 | µm |
| $w$ | 2 | µm |
| $t$ | 2 | µm |
| $L_t$ | 508 | µm |
| $w_t$ | 2 | µm |
| $t_t$ | 1.5 | µm |
| $P$ | 2320 | kg m$^{-3}$ |
| $v$ | .22 | |
| $I$ | 1.33 | µm$^4$ |
| $I_t$ | 1 | µm$^4$ |
| $m$ | 114 | ng |

Analytical expressions for the spring constants[4]:

$$k_{folded\ spring} = 24\frac{EI}{L^3} = 2\frac{Etw^3}{L^3} = 0.31\ \text{N/m}$$

$$k_{tether} = 12\frac{EI_t}{L_t^3} = \frac{Et_t w_t^3}{L_t^3} = 0.015\ \text{N/m}$$

$$k_{total} = 2(k_{folded\ spring} + k_{tether}) = 0.65\ \text{N/m} \tag{1}$$

From the finite element simulation we obtain (each component is simulated individually):

$$k_{folded\ spring} = 0.335\ \text{N/m}$$

$$k_{tether} = 0.016\ \text{N/m}$$

$$k_{total} = 2(k_{folded\ spring} + k_{tether}) = 0.702\ \text{N/m} \tag{2}$$

The values are in reasonable agreement considering the rounded edges that are not taken into account by the analytical solution and the finite mesh size that is used.

## 2.2  Electrical Properties of Comb Actuator

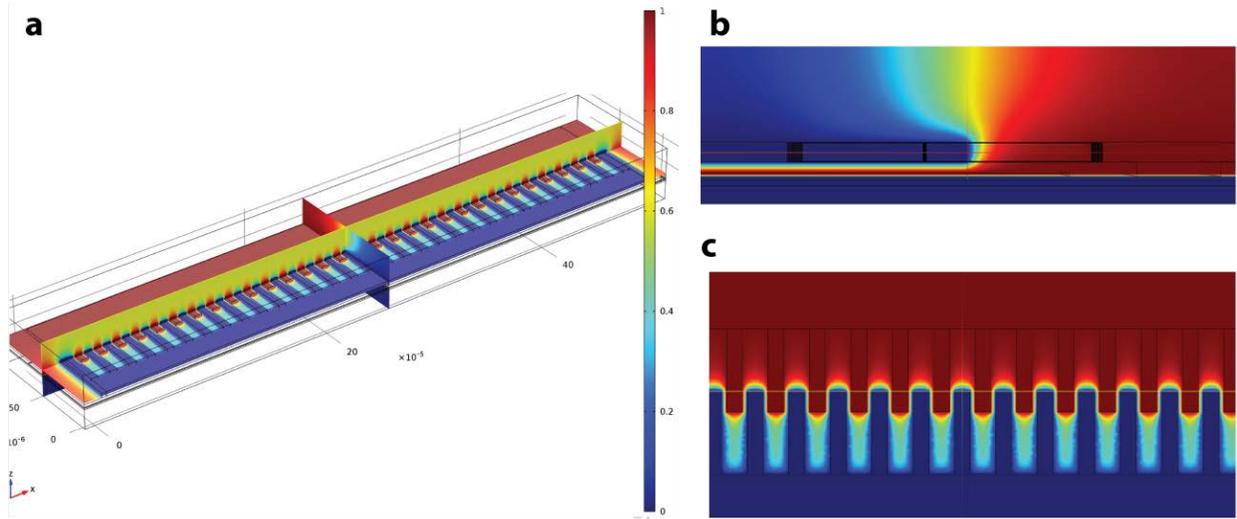

**Figure S-3. Finite Element Simulation of electrical properties of MEMS comb drive a**, Plot of potential for 1 V. **b,** side and **c,** top view of combs. The ground plane beneath the comb breaks the symmetry and results in an upward force[5].

The total capacitance due to the bottom plate is as high as $C$ = 7.91 pF. The capacitance of the combs themselves and the predicted change in capacitance due to mechanical motion is 2-3 orders of magnitude smaller respectively. This large stray capacitance makes it difficult to measure the change in capacitance due to small displacements.

The electrostatic force is given by:

$$F_i = -\frac{dE}{dx_i} = -\frac{V^2}{2}\frac{dC}{dx_i} \qquad (3)$$

Combining the Electromechancial properties of the device one obtains

$$F_i = -\frac{V^2}{2}C' = kx \qquad (4)$$

And hence the displacement is known for an applied voltage

$$x = -\frac{C'}{2k}V^2, \qquad (5)$$

This force is always attractive; hence the minus sign indicates a closing of the combs. Due to the geometry of the fingers such combs allow for a relatively large displacement compared to the gap size (the gap is 2-2.5 μm, the total displacement is almost 20 μm in each quadrant). This is possible because the motion is perpendicular to the gap. Hence the displacement does not change the value of $C'$. This is in stark contrast to parallel plate capacitors that not linear behavior and pull in occurs at 1/3 the gap size.

$\frac{C'}{2k}$ can be determined experimentally, here we determined $\frac{C'}{2k} = \frac{1}{787}\frac{\mu m}{V^2}$, From which $C' = 1.79 \frac{fF}{\mu m}$ is derived. For capacitive displacement measurement 1 nm resolution can be obtained if 1.8 aF change in capacitance can be detected. Commercial high sensitivity capacitive readout circuits are available with ~4 aF/Hz$^{1/2}$[6]. Such differential detection schemes could be used for active feedback control for dynamic control above the resonance frequency and reducing the ring-down time by imposing a critically damped electromechanical system.

We calibrated the device using the writing plate itself. Figure S4 depicts the relation between the applied voltage and displacement. For this graph the plate was moved in constant intervals of $V^2$ and at each step 12.0 nm of gold was deposited. The resulting SEM image (center of figure) was digitized and the absolute scale was calibrated to the SEM measurement of the maximum displacement. Figure S4 a shows excellent agreement with equation (5). From the linear fit the value $\frac{C'}{2k} = 1.270 \pm 0.001 \frac{nm}{V^2}$ can be extracted. Figure S4 b shows the residuals to the fit. The dotted lines illustrate that 80% of the measurements lie within 10 nm of the expected position. It is shown that the position of the plate is within 10 nm of the expected value with an 80% confidence level. It is believed that a significant part of the error results from the SEM image and the digitization process and is not due to the electromechanical response of the device itself. As is shown below the mechanical noise results in a vibration amplitude of ~65 nm, well above the actuation-displacement precision.

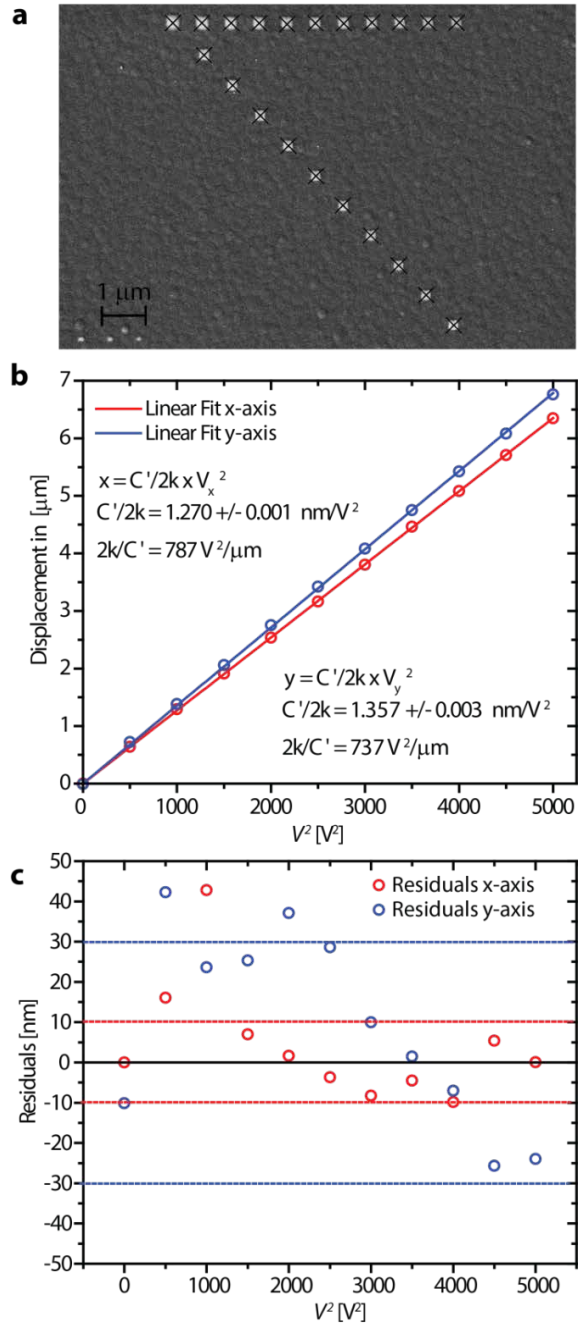

**Figure S-4. Displacement versus $V^2$ relation. a,** SEM image of gold dots spaced equally spaced in $V^2$. **b,** Displacement plotted against the voltage squared as determined from **a**, along the x and x-y axes. The linear fit is used to illustrate the electromechancial response $\frac{C_x`}{2k_x} = 1.270 \pm 0.001 \frac{nm}{V^2}$ and $\frac{C_y`}{2k_y} = 1.357 \pm 0.003 \frac{nm}{V^2}$ along the x and y axis respectively. **c,** Residuals plot of displacement versus $V^2$. 80% of the spread falls between the dotted lines corresponding to ±10 nm and ±30 nm for the x- and y-axis respectively. This is within the measurement accuracy.

## 3. Compressive Strain in the Tethers

Zygo Interferometry measurements show buckling at 0 V and 50 V actuation. Data is depicted in Figure S5.

In plane perpendicular forces cancel to zero in an ideal setup and can be neglected due to the high spring constant along that axis. The symmetry of the finger geometry along the z-axis is broken by a grounding plate beneath the comb. This results in a small positive out of plane force which disengages the fingers[5] and, due to the clamping of the springs, results in a downward tilt of the tethers. This would result in the central plate collapsing onto the substrate. However, residual compressive strain from the fabrication process buckles the tethers, lifting the aperture a safe distance above the silicon nitride (see Figure S5).

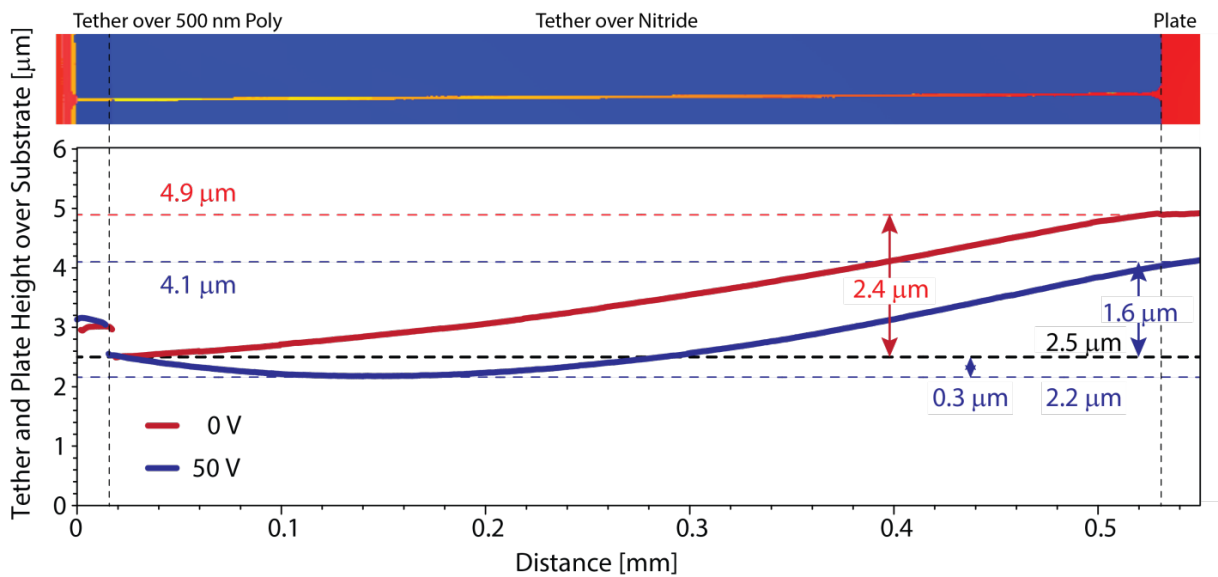

**Figure S-5. Interferometer measurements of tether profile.** At 0 V actuation the tethers are buckled up raising the central plate by 2.4 µm to 4.9 µm above the substrate (red trace). The actuated comb results in a force in the z axis. The folded springs act like a pivot and push the tether down. The blue trace shows how 50 V actuation lowers the plate by 0.8 µm, and the lowest point of the tether is only 2.2 µm above the substrate.

Buckling of the tethers increases geometric smearing by lifting the plate away from the substrate. Without buckling the out-of-plane forces and pivot would result in the plate touching the substrate. The resulting stiction severely reduces the displacement control and accuracy and can even make the aperture completely immobile. The buckling at 0 V is bi-stable and can occur away or towards the plate. After release, a micro-manipulator is used to poke the device and ensure the buckling is away from the substrate.

*4.     Experimental Setup*

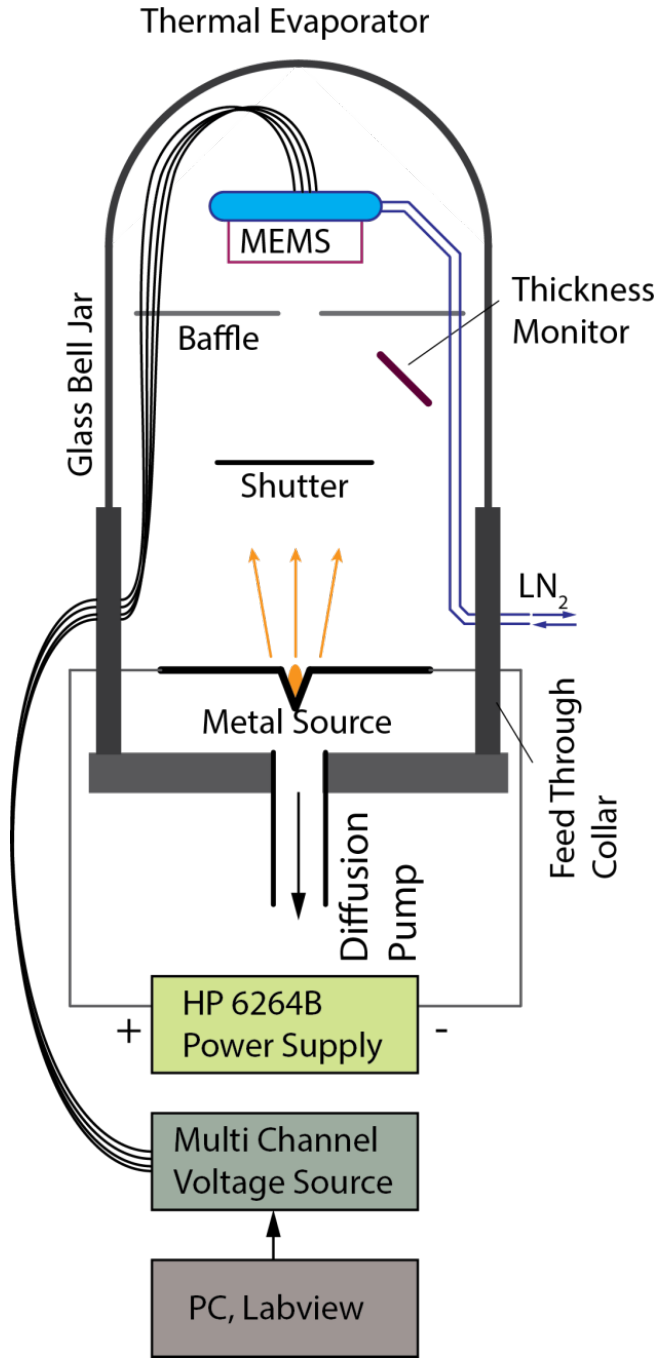

**Figure S-6. Experimental Setup.** Thermal Evaporator with MEMS voltage sources and computer controller. The baffle shields the setup from thermal radiation and unwanted gold deposition. A liquid nitrogen feed-through allows the stage to be cooled to 84 K. A thickness monitor measures the evaporation rate which is set by the dc power supply.

## 5. Dot Size

### 5.1 Geometric Smearing

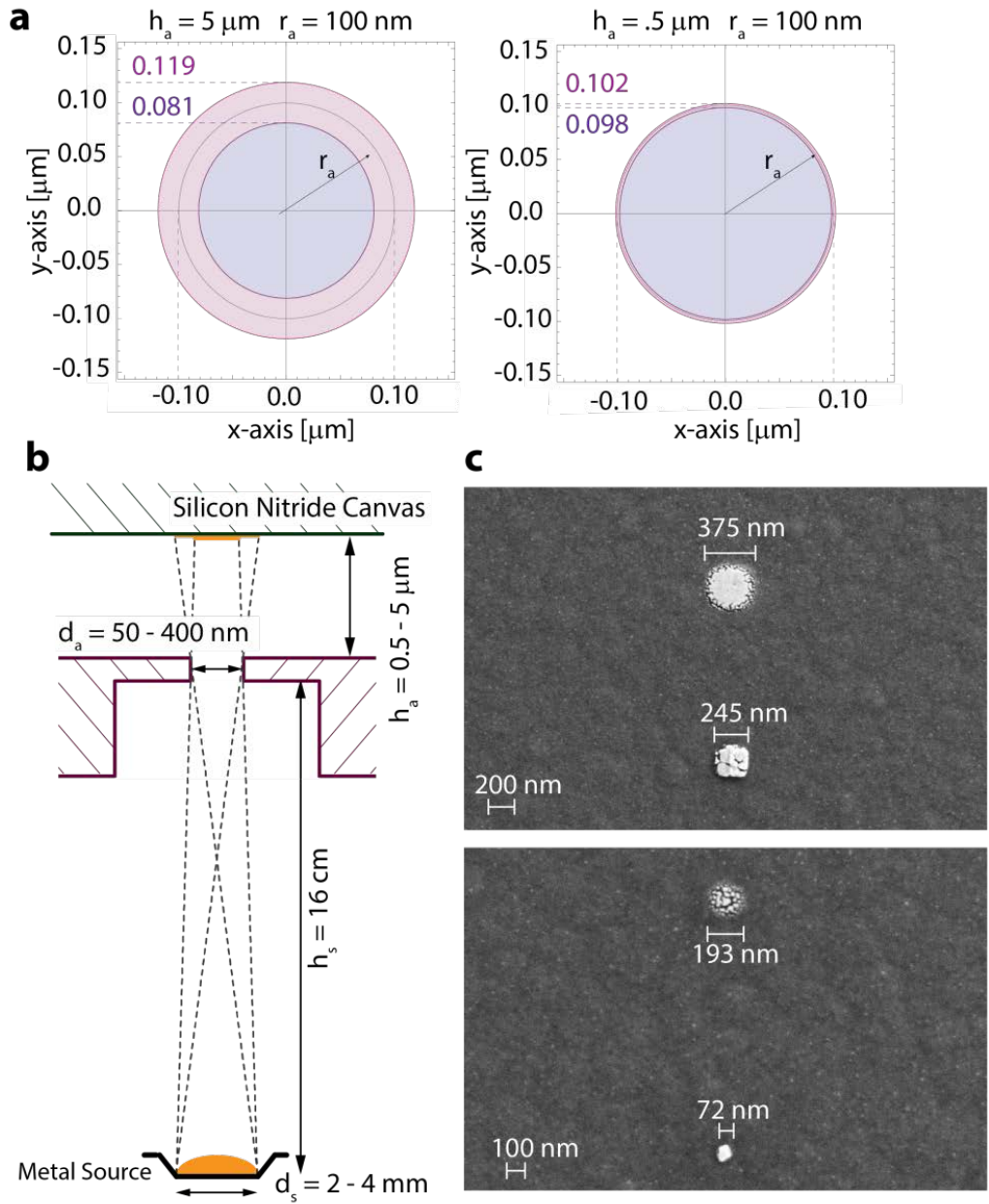

**Figure S-7. Calculation and images of geometric smearing. a,** Calculated smearing due to geometric effects for a 200 nm aperture and two aperture heights, 5.0 μm and 0.5 μm, for a source height of 20 cm. **b,** Finite source and aperture diameter as well as distance over the substrate results in a blurring of the traces, as is the case for stencil lithography. **c,** Experimental results for two aperture sizes. At (0,0) position the plate is ~5 μm above the substrate, the plate is moved and pulled down to the substrate. The resulting dots are ~121-130 nm smaller in diameter. The pull in is achieved electrostatically at moderate voltages (<10 V) as the plate forms a large capacitor with the doped silicon beneath the nitride.

## 5.2 Thermal Vibration Estimates

Intrinsic thermal noise estimates:

Spectral amplitude of damped driven harmonic oscillator

$$S_x(\omega) = \frac{1}{m^2} \frac{S_f(\omega)}{(\omega^2 - \omega_0^2)^2 + \left(\frac{\omega \omega_0}{Q}\right)^2} \tag{5}$$

Thermal spectral noise force:

$$S_{f_i}^{th} = \frac{4 k_B T m \omega_i}{Q} \tag{6}$$

On resonance

$$S_x^{th} = \frac{4 k_B T Q}{m \omega_0^3} \tag{7}$$

Resulting in an on resonance displacement of

$$\delta x_{rms} = \frac{Q S_{f_i}^{1/2}}{m \omega_0^2} \Delta \omega^{1/2}, \tag{8}$$

with $\Delta \omega^{1/2} = \sqrt{\frac{\omega}{Q}}$ being the resonant width. For this device $\delta x_{rms} = 0.27$ nm, using Q = 10000, $m = 4.5 \times 10^{-10}$ kg, $\omega_0 = 2\pi\, 10{,}000$ Hz, T = 300 K, and Boltzmann's constant $k_B$.

## 5.3 Discussion on Smearing

Geometric smearing accounts for a 60 nm widening of a stationary dot. Thermal motion will add less than 1 nm. Our measurements indicate smearing on the order of 125 nm. This indicates that mechanical vibration accounts for 65 nm widening. Adding vibration isolation will considerable reduce this external noise effect. Active feedback circuits can be used to further improve device resolution.

The depositions shown here are all resulting from slow plate movements with typical step sizes on the order of 1-20 nm, smaller than the expected geometric smearing. Vibration isolation, cooling to cryogenic temperatures and active feedback control circuits can dynamically reduce unwanted motion and ring-down, emulating a critically damped system[7].

*5.4   Smallest Dot*

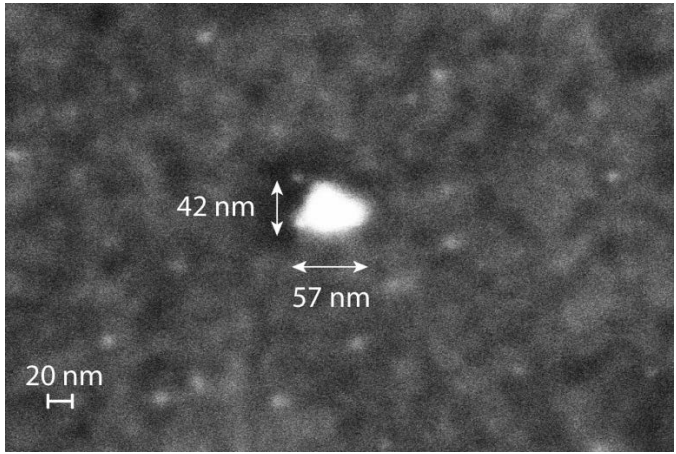

**Figure S-8. Smallest Aperture**. So far the smallest dot we have imaged is ~42 x 57 nm$^2$. Further optimizing the two step milling process using the FIB should result in apertures as small as 10 x10 nm$^2$.

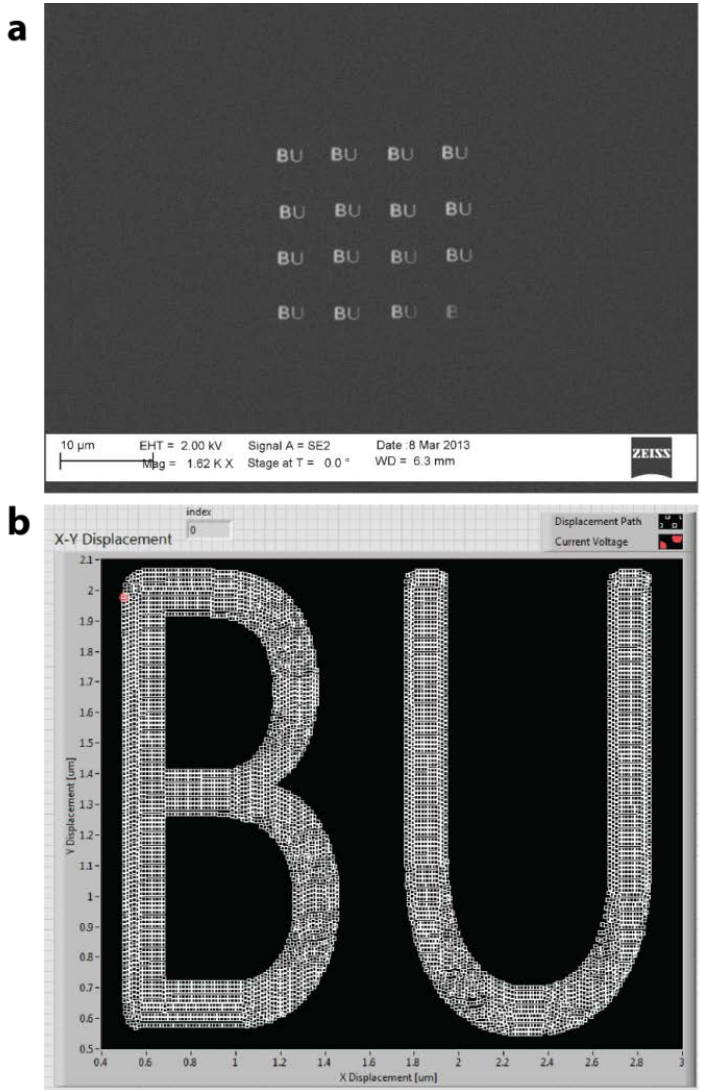

**Figure S-9. Writing an array using 16 apertures. a**, SEM micrograph of 16 BU patterns. **b**, Coordinates set by LabVIEW in μm.

Figure S8 a depicts an array fabricated out of 16 200 nm apertures. Figure S8 b shows the x-y coordinates in μm that LabVIEW converted to voltages. This illustrates how the plate can move like a conventional printer. The metal flux never stops, but the motion of the plate is so fast that there is virtually no metal accumulation outside the desired pattern.

## 7 Shutter Dynamics

The shutter is made of the same spring and comb actuator as each axis of the plate drivers. The mass is a bit higher due to the additional metal used to raise the shutter over the plate.

Considering the resonance frequency of this devices it can be opened and closed on resonance at ~6.5 kHz, resulting in a period of ~153 µs. For shorter intervals higher order modes could be actuated to eventually push the period below 50 µs.

Alternatively, it possible to drive the device above resonance by switching on a high voltage signal for a very short time. Consider the distance travelled in time t for an applied voltage V:

$$x = \frac{C`V^2}{4m} \frac{t^2}{(1+\frac{kt^2}{2m})}, \tag{9}$$

or the corresponding time needed to travel a distance x:

$$t = \sqrt{\frac{x}{\frac{C`V^2}{4m} - \frac{kx}{2m}}}, \tag{10}$$

Figure S9 a depicts the time needed to travel the distance x for three applied voltages.

A more general derivation is based on the damped harmonic oscillator. Equations 9 and 10 are close approximations for a critically damped system. These MEMS devices can have quality factors as high as $10^5$. Consider the second order inhomogeneous differential equation with damping parameter γ:

$$m\ddot{x} + \gamma\dot{x} + kx = F(t), \tag{11}$$

with

$$F(t) = 0 \ (t < 0)$$

$$F(t) = \frac{C`V^2}{2m} \ (t > 0). \tag{12}$$

The solution to the differential equation for *t>0* is

$$x(t) = \frac{C`V^2}{2mk}\left(1 - \frac{e^{-\frac{\gamma t}{2m}}}{\sin\varphi}\sin\left(\sqrt{1 - \frac{\gamma^2}{4m^2\omega_0^2}}\omega_0 t + \varphi\right)\right), \tag{13}$$

$$\cos\varphi = \frac{\gamma}{2m\omega_0}.$$

Equation (13) is plotted in Figure S9 b and c for three applied voltages.

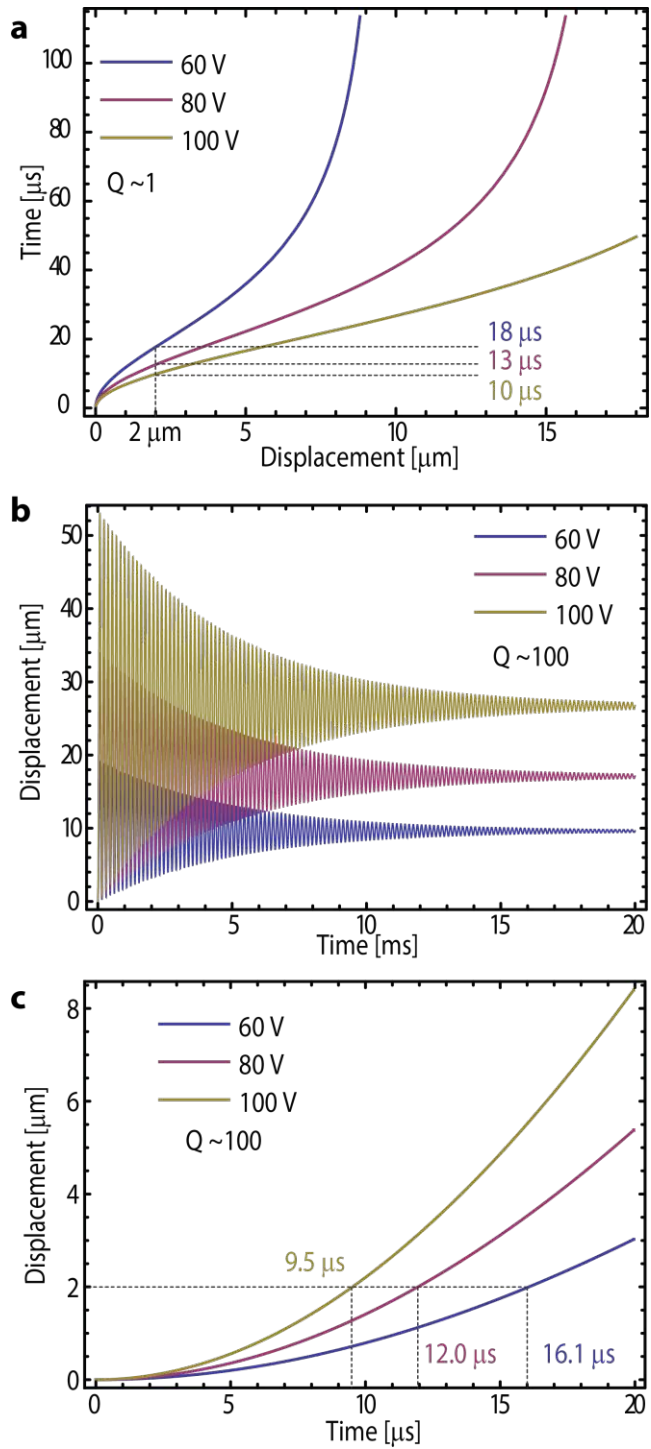

**Figure S-10. Displacement-Time theory plots. a,** Time needed to travel a set distance for $V_{SH}$ = 60 V, 80 V and 100 V. Units are µm and µs. This is valid for a critically damped resonator approximated by equation 10. **b**, displacement versus time for a damped harmonic oscillator with the voltages turned on at t=0. Q = 100, the relaxation time is order 10 ms. **c**, First 20 µs of plot b. The shutter is moved by 2 µm in under 10 µs for the largest applied voltage.

By applying 80 V, the shutter is moved by 2 µm in only 12 µs and for 120 V the opening (or closing time) falls to 8 µs. With only a single sided comb the structure will have to relax back into the zero position after the voltage is turned off. Again, for 2 µm actuation this would take ~49 µs. For higher quality factor devices the time is reduced, but not by much. High quality factors however will result in the aperture opening and closing again over the settling time which is on the order of 10 ms for a $Q = 100$ and can last as long as seconds for $Q = 10,000$. Where an unloaded MEMS devices of this type may well have a quality factor greater than 10,000, circuit loading resulting from the capacitor increases the dissipation proportionally to $V^2$. At 60 V the quality factor may well be <100. This exact value must still be determined experimentally. It must also be noted that the Keithely 2410 used has a setting time of 100 µs and hence sets a lower limit on the shutter speed.

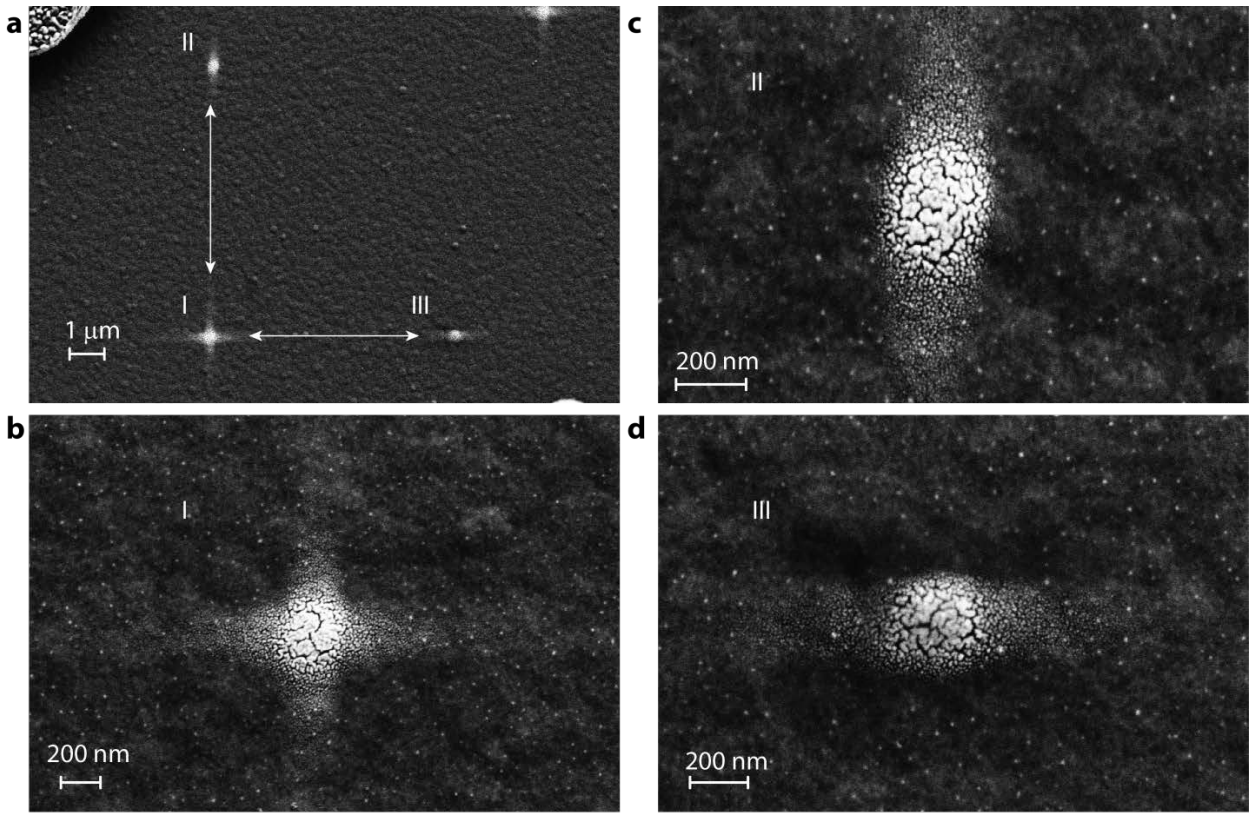

**Figure S-11. SEM Image of Ring-down**. **a**, Three dots defining the (0,0), (0,70) and (70,0)V position. Switching occurred in the sequence: I-II-I-III, with a wait time of 1 s at each point. **b**, **c** and **d** are close-up of the I, II and III positions respectively.

We can use deposition patterns to deduce information about the dynamics of our device. Figure S10 illustrates the actual ring-down response of the plate (expected to be very similar to the shutter). Metal is deposited while the plate moves between three positions, traveling a distance of 6.7 µm in the x direction or 7.5 µm along the y-axis. Each position is held for 1 s, which is expected to be much longer than the ring down time. The resulting deposited pattern illustrates how the plate overshoots. The thickness of the metal represents the envelope defined by the ring-down in Figure S9 b. It is essentially a time average of the displacement. If the deposited metal is sufficiently smooth, then this thickness could be measured using an AFM. This may allow us to extract the quality factor, which is a dynamic property, from the deposition trace.

Shortening the required distance results in shorter opening times, however, the relaxation time is constant. To reduce the closing time the mass can be lowered and the spring constant increased (shorter folded flexural springs and less gold). For additional dynamic control a second comb drive can be added to the shutter that can actively pull the shutter back. Currently the open-close time is a total of 12 µs + 49 µs = 61 µs. By simply adjusting the spring geometry this can be halved. Aligning the aperture with the shutter at half the maximal displacement the restoring force is greater and the relaxation time to the position before the aperture is opened will drop to the opening time. For the current setup it is therefore conceivable that opening and closing by 2 µm could be as short as 25 µs.

A double plate setup, where the shutter is controlled by additional springs and actuators, and aligning the shutter aperture within 100 nm to the writing aperture close to the will push the timing cycle into the µs range. Noting that both the plate and the shutter can move at speeds up to 1 m/s. Considering both plates moving over each other in opposite directions, the two 100 nm apertures would be open for only 50 ns, consistent with allowing the deposition of one or a few atoms.